\documentstyle[12pt]{article}
\newcommand{\nc}{\newcommand}
\nc{\renc}{\renewcommand}

\newcommand{\pv}{\mbox{\boldmath $p$}}
\newcommand{\kv}{\mbox{\boldmath $k$}}

\nc{\half}{{\textstyle{1\over2}}}

\nc{\etal}{\mbox{\it et al. }}
\nc{\ie}{{\it i.e.}}
\nc{\eg}{{\it e.g.}}

\renc{\thefootnote}{\arabic{footnote}}
\nc{\capt}[1]{{\bf Figure.} {\small\sl #1}}

\nc{\eqs}[2]{\mbox{Eqs.~(\ref{#1},\,\ref{#2})}}
\nc{\eq}[1]{\mbox{Eq.~(\ref{#1})}}

\nc{\figs}[2]{\mbox{Figs.~(\ref{#1},\,\ref{#2})}}
\nc{\fig}[1]{\mbox{Fig~.(\ref{#1})}}

\nc{\tag}[1]{\label{#1} \marginpar{{\footnotesize #1}}}
\nc{\mtag}[1]{\label{#1} \mbox{\marginpar{{\footnotesize #1}}}}
\renc{\baselinestretch}{1.2}
\newlength{\overeqskip}
\newlength{\undereqskip}
\nc{\be}[1]{\begin{equation} \mbox{$\label{#1}$}}
\nc{\bea}[1]{\begin{eqnarray} \mbox{$\label{#1}$}}
\nc{\Section}[2]{\section{#2}\label{#1}}
\nc{\Bibitem}[1]{\bibitem{#1}}
\nc{\Label}[1]{\label{#1}}
\nc{\eea}{\vspace{\undereqskip}\end{eqnarray}}
\nc{\ee}{\vspace{\undereqskip}\end{equation}}
\nc{\bdm}{\begin{displaymath}}
\nc{\edm}{\end{displaymath}}
\nc{\dpsty}{\displaystyle}
\nc{\bc}{\begin{center}}
\nc{\ec}{\end{center}}
\nc{\ba}{\begin{array}}
\nc{\ea}{\end{array}}
\nc{\bab}{\begin{abstract}}
\nc{\eab}{\end{abstract}}
\nc{\btab}{\begin{tabular}}
\nc{\etab}{\end{tabular}}
\nc{\bit}{\begin{itemize}}
\nc{\eit}{\end{itemize}}
\nc{\ben}{\begin{enumerate}}
\nc{\een}{\end{enumerate}}
\nc{\bfig}{\begin{figure}}
\nc{\efig}{\end{figure}}
\nc{\arreq}{&\!=\!&}
\nc{\arrmi}{&\!-\!&}
\nc{\arrpl}{&\!+\!&}
\nc{\arrap}{&\!\!\!\approx\!\!\!&}
\nc{\non}{\nonumber\\*}
\nc{\align}{\!\!\!\!\!\!\!\!&&}

\def\lsim{\; \raise0.3ex\hbox{$<$\kern-0.75em
      \raise-1.1ex\hbox{$\sim$}}\; }
\def\gsim{\; \raise0.3ex\hbox{$>$\kern-0.75em
      \raise-1.1ex\hbox{$\sim$}}\; }
\nc{\DOT}{\hspace{-0.08in}{\bf .}\hspace{0.1in}}
\nc{\Laada}{\hbox {$\sqcap$ \kern -1em $\sqcup$}}
\nc\loota{{\scriptstyle\sqcap\kern-0.55em\hbox{$\scriptstyle\sqcup$}}}
\nc\Loota{{\sqcap\kern-0.65em\hbox{$\sqcup$}}}
\nc\laada{\Loota}
\nc{\qed}{\hskip 3em \hbox{\BOX} \vskip 2ex}

\nc{\real}{{\rm I \! R}}
\nc{\Z}{{\sf Z \!\!\! Z}}
\nc{\complex}{{\rm C\!\!\! {\sf I}\,\,}}
\def\bigid{\leavevmode\hbox{\small1\kern-3.8pt\normalsize1}}
\def\id{\leavevmode\hbox{\small1\kern-3.3pt\normalsize1}}
\nc{\slask}{\!\!\!/}
\nc{\bis}{{\prime\prime}}
\nc{\pa}{\partial}
\nc{\na}{\nabla}
\nc{\ra}{\rangle}
\nc{\la}{\langle}
\nc{\goto}{\rightarrow}
\nc{\swap}{\leftrightarrow}

\nc{\EE}[1]{ \mbox{$\cdot10^{#1}$} }
\nc{\abs}[1]{\left|#1\right|}
\nc{\at}[2]{\left.#1\right|_{#2}}
\nc{\norm}[1]{\|#1\|}
\nc{\abscut}[2]{\Abs{#1}_{\scriptscriptstyle#2}}
\nc{\vek}[1]{{\rm\bf #1}}
\nc{\integral}[2]{\int\limits_{#1}^{#2}}
\nc{\inv}[1]{\frac{1}{#1}}
\nc{\dd}[2]{{{\partial #1}\over{\partial #2}}}
\nc{\ddd}[2]{{{{\partial}^2 #1}\over{\partial {#2}^2}}}
\nc{\dddd}[3]{{{{\partial}^2 #1}\over
        {\partial #2 \partial #3}}}
\nc{\dder}[2]{{{d #1}\over{d #2}}}
\nc{\ddder}[2]{{{d^2 #1}\over{d {#2}^2}}}
\nc{\dddder}[3]{{d^2 #1}\over
        {d #2 d #3}}
\nc{\dx}[1]{d\,^{#1}x}
\nc{\dy}[1]{d\,^{#1}y}
\nc{\dz}[1]{d\,^{#1}z}
\nc{\dl}[1]{\frac{d\,^{#1}l}{(2\pi)^{#1}}}
\nc{\dk}[1]{\frac{d\,^{#1}k}{(2\pi)^{#1}}}
\nc{\dq}[1]{\frac{d\,^{#1}q}{(2\pi)^{#1}}}

\nc{\cc}{\mbox{$c.c.$ }}
\nc{\hc}{\mbox{$h.c.$ }}
\nc{\cf}{cf.\ }
\nc{\erfc}{{\rm erfc}}
\nc{\Tr}{{\rm Tr\,}}
\nc{\tr}{{\rm tr\,}}
\nc{\pol}{{\rm pol}}
\nc{\sign}{{\rm sign}}
\nc{\bfT}{{\bf T }}

\def\GeV{{\rm\ GeV}}
\def\MeV{{\rm\ MeV}}
\def\keV{{\rm\ keV}}

\nc{\cA}{{\cal A}}
\nc{\cB}{{\cal B}}
\nc{\cD}{{\cal D}}
\nc{\cE}{{\cal E}}
\nc{\cG}{{\cal G}}
\nc{\cH}{{\cal H}}
\nc{\cL}{{\cal L}}
\nc{\cO}{{\cal O}}
\nc{\cT}{{\cal T}}
\nc{\cN}{{\cal N}}
\nc{\rvac}[1]{|{\cal O}#1\rangle}
\nc{\lvac}[1]{\langle{\cal O}#1|}
\nc{\rvacb}[1]{|{\cal O}_\beta #1\rangle}
\nc{\lvacb}[1]{\langle{\cal O}_\beta #1 |}
\nc{\bb}{\bar{\beta}}
\nc{\bt}{\tilde{\beta}}
\nc{\ctH}{\tilde{\cal H}}
\nc{\chH}{\hat{\cal H}}
\nc{\1}{\aa}
\nc{\2}{\"{a}}
\nc{\3}{\"{o}}
\nc{\4}{\AA}
\nc{\5}{\"{A}}
\nc{\6}{\"{O}}
\nc{\al}{\alpha}
\nc{\g}{\gamma}
\nc{\e}{\epsilon}
\nc{\eps}{\epsilon}
\nc{\lam}{\lambda}
\nc{\om}{\omega}
\nc{\Om}{\Omega}
\nc{\ve}{\varepsilon}
\nc{\mn}{{\mu\nu}}
\nc{\vp}{\varphi}

\nc{\advp}[3]{{\it  Adv.\ in\ Phys.\ }{{\bf #1} {(#2)} {#3}}}
\nc{\annp}[3]{{\it  Ann.\ Phys.\ (N.Y.)\ }{{\bf #1} {(#2)} {#3}}}
\nc{\apl}[3]{{\it  Appl. Phys. Lett. }{{\bf #1} {(#2)} {#3}}}
\nc{\apj}[3]{{\it  Ap.\ J.\ }{{\bf #1} {(#2)} {#3}}}
\nc{\apjl}[3]{{\it  Ap.\ J.\ Lett.\ }{{\bf #1} {(#2)} {#3}}}
\nc{\app}[3]{{\it Astropart.\ Phys.\ }{{\bf #1} {(#2)} {#3}}}
\nc{\cmp}[3]{{\it  Comm.\ Math.\ Phys.\ }{{ \bf #1} {(#2)} {#3}}}
\nc{\cqg}[3]{{\it  Class.\ Quant.\ Grav.\ }{{\bf #1} {(#2)} {#3}}}
\nc{\epl}[3]{{\it  Europhys.\ Lett.\ }{{\bf #1} {(#2)} {#3}}}
\nc{\ijmp}[3]{{\it Int.\ J.\ Mod.\ Phys.\ }{{\bf #1} {(#2)} {#3}}}
\nc{\ijtp}[3]{{\it Int.\ J.\ Theor.\ Phys.\ }{{\bf #1} {(#2)} {#3}}}
\nc{\jmp}[3]{{\it  J.\ Math.\ Phys.\ }{{ \bf #1} {(#2)} {#3}}}
\nc{\jpa}[3]{{\it  J.\ Phys.\ A\ }{{\bf #1} {(#2)} {#3}}}
\nc{\jpc}[3]{{\it  J.\ Phys.\ C\ }{{\bf #1} {(#2)} {#3}}}
\nc{\jap}[3]{{\it J.\ Appl.\ Phys.\ }{{\bf #1} {(#2)} {#3}}}
\nc{\jpsj}[3]{{\it J.\ Phys.\ Soc.\ Japan\ }{{\bf #1} {(#2)} {#3}}}
\nc{\lmp}[3]{{\it Lett.\ Math.\ Phys.\ }{{\bf #1} {(#2)} {#3}}}
\nc{\mpl}[3]{{\it  Mod.\ Phys.\ Lett.\ }{{\bf #1} {(#2)} {#3}}}
\nc{\ncim}[3]{{\it  Nuov.\ Cim.\ }{{\bf #1} {(#2)} {#3}}}
\nc{\np}[3]{{\it  Nucl.\ Phys.\ }{{\bf #1} {(#2)} {#3}}}
\nc{\pr}[3]{{\it Phys.\ Rev.\ }{{\bf #1} {(#2)} {#3}}}
\nc{\pra}[3]{{\it  Phys.\ Rev.\ A\ }{{\bf #1} {(#2)} {#3}}}
\nc{\prb}[3]{{\it  Phys.\ Rev.\ B\ }{{{\bf #1} {(#2)} {#3}}}}
\nc{\prc}[3]{{\it  Phys.\ Rev.\ C\ }{{\bf #1} {(#2)} {#3}}}
\nc{\prd}[3]{{\it  Phys.\ Rev.\ D\ }{{\bf #1} {(#2)} {#3}}}
\nc{\prl}[3]{{\it Phys.\ Rev.\ Lett.\ }{{\bf #1} {(#2)} {#3}}}
\nc{\pl}[3]{{\it  Phys.\ Lett.\ }{{\bf #1} {(#2)} {#3}}}
\nc{\prep}[3]{{\it Phys\. Rep.\ }{{\bf #1} {(#2)} {#3}}}
\nc{\prsl}[3]{{\it Proc.\ R.\ Soc.\ London\ }{{\bf #1} {(#2)} {#3}}}
\nc{\ptp}[3]{{\it  Prog.\ Theor.\ Phys.\ }{{\bf #1} {(#2)} {#3}}}
\nc{\ptps}[3]{{\it  Prog\ Theor.\ Phys.\ suppl.\ }{{\bf #1} {(#2)} {#3}}}
\nc{\physa}[3]{{\it  Physica\ A\ }{{\bf #1} {(#2)} {#3}}}
\nc{\physb}[3]{{\it  Physica\ B\ }{{\bf #1} {(#2)} {#3}}}
\nc{\phys}[3]{{\it Physica\ }{{\bf #1} {(#2)} {#3}}}
\nc{\rmp}[3]{{\it  Rev.\ Mod.\ Phys.\ }{{\bf #1} {(#2)} {#3}}}
\nc{\rpp}[3]{{\it Rep.\ Prog.\ Phys.\ }{{\bf #1} {(#2)} {#3}}}
\nc{\sjnp}[3]{{\it Sov.\ J.\ Nucl.\ Phys.\ }{{\bf #1} {(#2)} {#3}}}
\nc{\spjetp}[3]{{\it Sov.\ Phys.\ JETP\ }{{\bf #1} {(#2)} {#3}}}
\nc{\yf}[3]{{\it Yad.\ Fiz.\ }{{\bf #1} {(#2)} {#3}}}
\nc{\zetp}[3]{{\it Zh.\ Eksp.\ Teor.\ Fiz.\  }{{\bf #1}  {(#2)} {#3}}}
\nc{\zp}[3]{{\it Z.\ Phys.\ }{{\bf #1} {(#2)} {#3}}}
\nc{\ibid}[3]{{\sl ibid.\ }{{\bf #1} {#2} {#3}}}
\nc{\rf}[1]{(\ref{#1})}
\nc{\nn}{\nonumber \\*}
\nc{\bfB}{\bf{B}}
\nc{\bfv}{\bf{v}}
\nc{\bfx}{\bf{x}}
\nc{\bfy}{\bf{y}}
\nc{\vx}{\vec{x}}
\nc{\vy}{\vec{y}}
\nc{\oB}{\overline{B}}
\nc{\oI}{\overline{I}}
\nc{\oR}{\overline{R}}
\nc{\rar}{\rightarrow}
\nc{\ti}{\times}
\nc{\slsh}{\hskip-5pt/}
\nc{\sm}{Standard~Model~}
\nc{\MP}{M_{\rm Pl}}
\nc{\tp}{t_{\rm Pl}}
\nc{\ave}{\bar{E}}

\nc{\eff}{{\rm eff}}
\nc{\kk}{\vek{k}}
\nc{\pp}{{\rm p}}
\nc{\ga}{g_{a\gamma}}
\nc{\vv}{\\}
\nc{\eee}{{\bf E}}
\nc{\bbb}{{\bf B}}
\nc{\qcd}{T_{\rm QCD}}
\nc{\G}{\rm \ G}
\def\vec#1{{\bf #1}}
\begin{document}
{\title{\vskip-2truecm{\hfill {{\small HU-TFT-96-34\\
        \hfill hep-ph/9608354\\
        }}\vskip 1truecm}
{\bf Cosmological abundances of right-handed neutrinos}}

%\vspace{1.2cm}

{\author{
{\sc Kari Enqvist$^{1}$, Petteri Ker\"anen$^{2}$, Jukka Maalampi$^{3}$
}\\
and \\
{\sc Hannes Uibo$^{4}$}\\
{\sl\small Department of Physics, P.O. Box 9,
FIN-00014 University of Helsinki,
Finland}
}
{\date{17 August 1996}}
\maketitle
\vspace{2cm}
\begin{abstract}
\noindent
We study the equilibration of the right-helicity states of light
Dirac neutrinos in the early universe by solving the momentum dependent
Boltzmann equations numerically. We show that the main effect is due
to electroweak gauge boson poles, which enhance thermalization rates by some
three orders of magnitude. The right-helicity states of tau neutrinos
will be brought in equilibrium independently of their initial distribution
at a temperature above the poles if $m_{\nu_\tau}\gsim 10~\keV$.
\end{abstract}
\vfil
\footnoterule
{\small $^1$enqvist@rock.helsinki.fi; \vskip-1pt\noindent
$^{2}$keranen@phcu.helsinki.fi;\vskip-1pt\noindent
$^{3}$maalampi@phcu.helsinki.fi;\vskip-1pt\noindent
$^{4}$ulbo@phcu.helsinki.fi (on a leave of
absence from Institute of Physics, Tartu, Estonia)}
\thispagestyle{empty}
\newpage
\setcounter{page}{1}

%%%%%%%%%%%%%%%%%%%%%%%%%%%%%%%%%%%%%%%%%%%%%%%%%%%%%%%%%%%%%%%%%%%%%%%%%%%%%%
\section{Introduction}
%%%%%%%%%%%%%%%%%%%%%%%%%%%%%%%%%%%%%%%%%%%%%%%%%%%%%%%%%%%%%%%%%%%%%%%%%%%%%%
Primordial nucleosynthesis is a remarkable probe of neutrino properties
\cite{subir}. Although recently the increasing accuracy of the cosmological
data, such as observations related to the primordial
abundances of  helium, deuterium and the other light elements,
has emphasized  systematic errors inherent to primordial nucleosynthesis
analysis, there remains
a great potential for constraining neutrino physics via cosmological
observations. To some extent
primordial nucleosynthesis could be sensitive even
to the Dirac vs. Majorana  nature of neutrinos, because in the Dirac case
the small relic abundance of the inert right-handed component of
Dirac neutrino would also contribute at nucleosynthesis. Of course,
presently one cannot  hope to  differentiate between the
Dirac and Majorana nature of neutrinos on cosmological grounds, but
in principle this is an interesting problem. In practise,
because of the smallness of the electron neutrino mass, from this point of view
only the right-handed components of
$\nu_\mu$ and $\nu_\tau$ can have interesting relic abundances.

Naively, the cosmological density of light right-handed
neutrinos $\nu_R$ (or rather the right-helicity states of light
neutrinos, $\nu_+$) is expected to be very
small. If $\nu_R$'s ever were  in equilibrium, they decoupled very early
because of their low capability of interacting. Assuming this took place well
above the
electroweak phase transition temperature $T_{\rm EW}$, at the onset of
primordial nucleosynthesis
the contribution of the right-handed component of a Dirac neutrino
is given by $\rho_R\simeq [g_*(1\MeV)/g_*(T_{\rm EW})]^{4/3}\rho_L\simeq
0.044\rho_L$, where $g_*(T)$ is the effective number of degrees of freedom in
thermal
equilibrium in the temperature $T$ and
$\rho_L$ is the equilibrium energy density of
 left-handed neutrino. This assumes that at
high temperature we may consider the gas of quarks, leptons, and gauge bosons
nearly ideal,
which may not be true. A recent lattice simulation
\cite{cond} of the QCD energy density above the
critical temperature has revealed that the actual energy density is
some 15\% smaller than expected, which might signify the existence
of a condensate at high $T$. One also assumes that there is no
significant entropy production either at the electroweak or QCD
phase transitions. Lattice simulations seem to indicate that
this is true for QCD, and the latent heat in the electroweak phase
transition is also known to be small \cite{eikr}.

The right-helicity states of Dirac neutrinos are not
completely inert in the Standard Model \cite{gaemers} but can be
produced (and destroyed) in spin-flip transitions induced by the Dirac mass
\cite{mass,kimmo} or the neutrino magnetic moment.
If the neutrino mass is large enough, $\nu_+$ would be produced in
collisions below the QCD phase transition temperature, resulting effectively
in an additional neutrino species at nucleosynthesis.  An important
source of $\nu_+$'s are also the non-equilibrium neutrino scatterings and
decays of pions, as
was pointed out in \cite{kimmo}. This gives rise to the  bound
$m_{\nu_{\mu}} \lsim 130 \keV$ and $m_{\nu_{\tau}} \lsim 150 \keV$
\cite{FieldsKimmoOliivi}, using $T_{\rm QCD}=100$~MeV and assuming
that  nucleosynthesis allows  less than $0.3$ extra neutrino
families. This is very restrictive bound for tau neutrinos since, basing
on primordial nucleosynthesis arguments,
there seems to be no window of opportunity for a sufficiently stable ($\tau_\nu
\gsim
10^2$~sec) tau neutrino
in the MeV region because of the production of non-equilibrium electron
neutrinos in $\nu_\tau\bar{\nu}_\tau$ annihilations
\cite{DolgovPastorValle}.

The actual cosmological density of the right-helicity neutrinos depends
not only on the production rate near the QCD phase transition, but also
on whether the right-helicity neutrinos had a chance to equilibrate
at some point during the course of the evolution of the universe.
%In the Standard Model,
%assuming no new interactions, one would tend to argue only interactions
%close to the QCD phase transition are important in this respect.
In this paper we wish to point out that at $T\lsim 100 \GeV$
there is an enhancement of the $\nu_+$ production rate due to the electroweak
gauge
boson poles.\footnote{ The question of the cosmic abundance of the right-handed
component of
Dirac neutrinos was previously studied in  \cite{Shapiro}. There
the pole effect was not considered.}
As a consequence there is a temperature range  in which the production rate can
exceed the
expansion rate of the universe and the right-helicity neutrinos may be brought
into thermal equilibrium with other light  particles. This will take place
if the
neutrino mass is sufficiently large. We show that for the tau neutrino
the mass
limit is  about 10
keV, and  of the same order of magnitude for the muon neutrino.  At the
temperature $T=1$ MeV the energy density of right-helicity
tau neutrinos with a mass of 10 keV is found to be about 6\% of the energy
density  of an ordinary
left-helicity neutrinos, i.e. their contribution to the effective number of
neutrino species is
$\Delta {N}_{\nu} \simeq 0.06$. The contribution of right-helicity neutrinos
with a smaller mass
depends, apart from the mass, on their initial energy density above the
electroweak scale.

The paper is organized as follows. In Section 2 we will list all
tree-level
spin-flip reactions  where right-helicity neutrinos are
produced and discuss their relative importance. We will also demonstrate
the $W$-pole effect
by considering  the reactions $u\overline d\to\nu_+\tau^+$
and $\tau^-u\to\nu_+d$ as an example for  the right-helicity tau
neutrino production. In Section 3 we consider the evolution of right-handed
neutrino density
and describe the numerical method  used in solving the Boltzman equation. The
results and a
discussion is presented in Section 4.

%%%%%%%%%%%%%%%%%%%%%%%%%%%%%%%%%%%%%%%%%%%%%%%%%%%%%%%%%%%%%%%%%%%%%%%%%%%%%%
\section{Right-helicity neutrino production}
%%%%%%%%%%%%%%%%%%%%%%%%%%%%%%%%%%%%%%%%%%%%%%%%%%%%%%%%%%%%%%%%%%%%%%%%%%%%%%
\subsection{Processes}
We shall consider the production of right-helicity neutrinos in the
Standard Model without assuming any new interactions except the Yukawa
interactions of the
right-handed neutrino with the  scalar doublet. The Yukawa coupling is the
origin of
Dirac neutrino mass and provides a spin-flip operator responsible for the
interactions of the
right-helicity neutrinos, the probability of the spin flip being proportional
to the
neutrino mass squared. Another source for the spin-flip is neutrino magnetic
moment
\cite{Fujikawa}, which appears in the Standard Model first at one-loop
level and is, therefore, too small to be of any significance for our
considerations.

We shall consider light neutrinos with mass less than 0.2 MeV at temperatures
between 1 MeV and 100 GeV. Since each right-helicity neutrino introduces in the
matrix
element a factor $m_{\nu}/|\pv_{\nu}|$, we may safely neglect processes in
which more than
one right-helicity neutrino is involved. All relevant processes are listed
(up to crossing)
in Table 1.

There are  68 purely fermionic
$2\to 2$ processes  in which a right-helicity
muon or tau neutrino can be produced. In addition,
a right-helicity tau neutrino can also be produced in  11 lepton and quark
three-body
decays, and the  muon neutrino in another set of 11 three-body
decays. Since we are especially interested in interactions occuring around
the poles of weak gauge bosons, we have to consider also processes involving
$W^{\pm}$, $Z$
and $H$. There are
 16  such processes.
Finally, there are 3 two-body decays  of $W^{\pm}$, $Z$
and $H$ bosons which are capable producing right-helicity muon and tau
neutrinos. However, as we will show below,
the processes involving gauge or Higgs bosons can  be neglected in comparision
with the
purely fermionic processes.

In what follows, for definiteness, we will consider only processes
including right-helicity tau neutrinos. Since we are mainly interested
on processes at temperatures above the muon and tau lepton masses, we
expect that the results below are roughly valid also for muon neutrinos.

\subsection{Production rates}

Let us consider a $2\to 2$ scattering $a+b\to \nu_{+}+d$ where one of the
final state particles is a right-helicity neutrino.
To estimate the relative importance of various processes we approximate
the thermally averaged production rate per one $\nu_+$ by
\bea{rate}
\Gamma_+ & = &  \frac{1}{n^{\rm FD}_{+}}
           \int
           d\Pi_a d\Pi_b d\Pi_+ d\Pi_d (2\pi )^4
           \delta^{(4)} (p_a + p_b - p_+ - p_d)
               S |{\cal M}_{ab\to +d}|^2 \nonumber \\
           & & \quad\quad\quad\quad \times f_a^{\rm FD} f_b^{\rm FD}
               (1-f_+^{\rm FD}) (1-f_d^{\rm FD})~,
\eea
where $n^{\rm FD}_{+}$ is the the equilibrium
number density of the right-handed neutrinos, $d\Pi_i \equiv d^3p_i/((2\pi)^3
2E_i)$,  $S$
is the symmetry factor taking into account identical particles in the initial
and/or final
states, and $f_i^{\rm FD}$ are Fermi-Dirac distribution functions.

At high $T$ the rate \eq{rate} is infrared sensitive to the thermal
corrections in the propagators. In general, the structure of the gauge
boson propagators at finite $T$ is complicated because Lorentz
symmetry is lost, which results in separate transverse and longitudinal
self energies $\Pi_T(\omega, \kv)$ and $\Pi_L(\omega, \kv)$.
In practise, however, the
leading thermal effect arises from small momenta, so that in most cases
it is an excellent approximation just to modify the propagators by
introducing a Debye mass $M^2(T)=\Pi_L(\omega, \kv=0)$, which we approximate by
$M_i^2(T)\simeq M_i^2 + 0.1\; T^2$ ($i=W,Z$). (This modification is necessary
in  t-channel propagators
only, since s-channel propagators are not infrared sensitive.)
In this approximation external particles  or interaction vertices
do not receive thermal corrections.
Thus the dispersion relations in the thermal distributions
in \eq{rate} remain unchanged. Admittedly, this is a simplistic approach, but
for our purposes, and for the  desired accuracies, this should be sufficient.
In fact,  in the region of interest the effects due to
thermal masses turn out to be very small.

In addition to  thermal corrections, we must account for the
imaginary parts of the gauge boson propagators, or the widths.
This is particularly important for the s-channel.
Thermal corrections will generate additional imaginary
parts both in the s-channel and  t-channel, but at $T\lsim 100$ GeV
they may safely be neglected.

A technical detail worth pointing out is that for a fixed helicity,
the spin-flip matrix elements are not Lorentz-invariant since the direction of
the
spin picks out a preferred frame of reference, as was emphasized in
\cite{kimmo}. Indeed, a Lorentz boost
changes the helicity of the particle, so that sometimes a fixed-helicity
reaction forbidden
in the CM may actually take place in another frame.
This means that it is not sufficient to compute matrix
elements just in e.g. the CM-frame, but instead one should a use a general
frame.

The main purpose of this  paper is to show that interactions at the
weak boson pole may
bring the right-helicity neutrinos to thermal equilibrium. To demonstrate
this effect, let us
consider  the the t-channel reaction $\tau^- u\to\nu_{+} d$ and its
crossed s-channel reaction $u\overline d\to\nu_{+}\tau^+$, where $\nu_+$ is a
right
helicity tau neutrino. The matrix element for the t-channel process reads
\be{Mt}
{\cal M}  = \frac{G_F}{\sqrt{2}} V_{ud}R_W (q^2 )
       (-g_{\mu \nu} + \frac{ q_{\mu} q_{\nu} }{M_W^2 } )
         \:
      \bar{u}_{\nu} \gamma^{\mu} (1 - \gamma_5 ) u_{\tau}
      \bar{u}_d \gamma^{\nu} (1- \gamma_5 ) u_u~,
\ee
where $q=p_{\tau}-p_{\nu}$, $V_{ud}$ is the appropriate CKM matrix element, and
as in s-channel, the propagator gives rise to a term of the form
\bea{RW(q)}
R_W (q^2 ) = \frac{M_W^2}{q^2 - M_W^2(T) + i\Gamma_W M_W }~.
\eea
As mentioned above, the width $\Gamma_W$ is important only for the
s-channel. Note the $T$-dependence in the numerator.
The matrix element squared summed over spins of $\tau^-$ and quarks is then
\bea{totele}
|{\cal M}|^2 =
G_F^2|V_{ud}|^2|R_W (q^2)|^2 (T_{gg} + T_{gq} + T_{qq})~,
\eea
where
\bea{Ttensors}
T_{gg} & = & 64 (p_{\tau} \cdot p_{u} ) (K_{\nu} \cdot p_{d} )~,
    \nonumber \\ [2 mm]
T_{gq} & = & \frac{32}{M_W^2 }
              \left\{ \left[ m_{u}^2 (p_{\nu} \cdot p_{d} )
             - m_{d}^2 (p_\nu \cdot p_u )\right] (K_{\nu} \cdot p_{\tau} )
\right.
    \nonumber \\ [2 mm]
           & & \: -\,
             \left[ (p_{\tau} \cdot p_{\nu} ) - m_{\tau}^2 \right]
             \left[ m_{u}^2 (K_{\nu} \cdot p_d) - m_d^2 (K_{\nu}\cdot p_{u} )
\right]
    \nonumber \\ [2 mm]
           & & \: \left.
             -\: m_{\nu}^2 \left[ m_{u}^2 (p_{\tau} \cdot p_d ) -
               m_d^2 (p_{\tau}\cdot p_u   ) \right] \right\}~,
    \nonumber \\ [2 mm]
T_{qq} & = & \frac{16}{M_W^4 } \left\{ (m_{\tau}^2 - m_{\nu}^2 )  (K_{\nu}\cdot
p_{\tau})
             + 2m_{\nu}^2 \left[(p_{\tau} \cdot p_{\nu} ) - m_\tau^2 \right]
\right\}
    \nonumber \\ [2 mm]
           & & \: \times \,
             \left\{\left[ m_{u}^2
             + m_d^2\right] (p_{u} \cdot p_d ) - 2 m_{u}^2 m_d^2 \right\}~.
\eea
Here $K^\lambda\equiv p^{\lambda} -m s^{\lambda}$, with the
spin four-vector $s^{\lambda}$  for particles with a definite helicity $h$
given by
\bea{spinvector}
s^\lambda = h \, \left( \frac{|\pv|}{m},
                 \frac{E}{m}\frac{\pv}{|\pv|} \right) ~.
\eea
In the ultra-relativistic limit $K^\lambda$ can be approximated as
\bea{ultra}
K^{\lambda} &\simeq & 2p^\lambda\:\:\: {\rm for}\: h=-1~,\label{LH-K}\\
K^{\lambda} &\simeq & \frac{m^2}{2|\pv|^2} (|\pv|,-\pv )
            \:\:\: {\rm for}\: h=+1~.
\label{RH-K}
\eea
Accordingly  the quantities $T_{gg}$, $T_{gq}$ and
$T_{qq}$ in
\eq{Ttensors} have in the case of  $\nu_+$ production typical sizes
given by
\bea{Ttensorprop}
T_{gg} & \sim& m^2_\nu T^2~, \cr
T_{gq} & \sim& ({ m_q }/{ M_W })^2 m^2_\nu T^2~, \cr
T_{qq} & \sim& ({ m_q }/{ M_W^2 })^2 m^2_\nu T^4~.
\eea
For the production of $\nu_-$ one has  $ T_{gg} \sim
T^4~$,
$ T_{gq} \sim ({ m_q }/{ M_W })^2 T^4~$ and
$T_{qq} \sim ({ m_q m_\tau }/{ M_W^2 })^2  T^4~$.
For both $\nu_+$ and $\nu_-$ at $T\lsim M_W$ and in the case of quarks other
than the top,
the terms $T_{gq}$ and $T_{qq}$ can be neglected in comparison with $T_{gg}$.
The rates
for $\nu_+$ are suppressed by a factor of the order of $m_{\nu}^2/T^2$ compared
with those
for $\nu_-$.  However, at higher temperatures, $T\gg M_W$, the $T_{qq}$ term
starts to
dominate the production rate of $\nu_+$ because of terms not proportional to
$K_\nu$.
Consequently the production rates of $\nu_+$ are in this case suppressed by a
{\it constant}
factor of the order of $m_q^2m_{\nu}^2/M_W^4$.

The s-channel matrix element can easily be obtained from \eq{totele}
by crossing. The thermal rates for both s- and t-channel processes can
then be found from \eq{rate}. A numerical integration results in
the curves displayed in Fig. 1. Here we have for simplicity ignored
the small finite temperature effects.
It can be see from the figure that the effect of the pole is spread over a
relatively large
temperature  range. This is  a consequence
of thermal averaging. Nevertheless, the enhancement in the s-channel is
apparent.  The results
presented in Fig. 1 are for the tau neutrino of mass $20\keV$,
but they are equally valid for the muon neutrino.
In this  case the thermally averaged production rate $\Gamma_+$ of the
s-channel process is
seen to exceed  in a certain temperature interval the expansion rate of the
universe, given by the
Hubble parameter
\be{hubble}
H\equiv{\dot R\over R}=\left({8\pi\rho_{\rm tot}\over 3M_{Pl}^2}\right)^{1/2}~.
\ee
 Hence, with this reaction alone  a 20 keV right-helicity
tau (and muon) neutrino would be brought into  thermal equilibrium while
universe  cools through this stage.  The complete analysis described in
the next
section, which is based on solving the Boltzmann equation with all the
relevant processes included,
confirms this expectation. The pole effects are indeed important
for an estimate of the relic density of right-helicity neutrinos.

In  addition to fermionic $2\to 2$ scattering processes, at high $T$
a potential source for $\nu_+$'s are $2\to 2$ processes that involve
the gauge or the Higgs bosons in the final or initial state (we dub such
processes  ``bosonic'').
Let us consider as
an example the process $\tau^-\gamma \rightarrow \nu_{+} W^-$, where the photon
couples
either to the charged lepton (Compton scattering) or to the W-boson in a
three-boson vertex. For comparision, the thermally averaged rate of this
process is  also  displayed in Fig. 1.
One can see that it can be
neglected even at temperatures around the pole, because there the production
rate of
the s-channel purely fermionic process is about three orders of magnitude
higher.

We have not considered all possible bosonic processes.
It is however very plausible that generically among
the $2\to 2$ scattering processes  only the purely fermionic processes are
important,
and among these, s-channel dominates over t-channel because of the pole in the
s-channel.
In what follows, we will always disregard bosonic processes.

The importance of decays is less straightforward to discern.
Numerical inspection reveals that the two-body decays listed in
the Table 1 can be neglected: their contribution to the total rate is
$\sim {\cal O}(10^{-11})$ at $T\gsim 10$~GeV, at lower temperatures
their contribution vanishes exponentially.
Three-body decays are more important. We find that while the total
contribution from three-body decays is negligible at higher momenta,
$|{\bf p}_+|/T \gsim 3$, it is as large as few per cents for
$|{\bf p}_+|/T \sim 3$ at $T \lsim 1$~GeV, increasing up to $\sim 30\%$
for very small momenta. For such small momenta also
$\nu_+$ production by $t\to b\tau^+\nu_+$,which has its maximal contribution
($\sim 10\%$) around $T\sim 30$~GeV, is important.

%%%%%%%%%%%%%%%%%%%%%%%%%%%%%%%%%%%%%%%%%%%%%%%%%%%%%%%%%%%%%%%%%%%%%%%%%%%%%%
\section{Evolution of the right-helicity neutrino density}
%%%%%%%%%%%%%%%%%%%%%%%%%%%%%%%%%%%%%%%%%%%%%%%%%%%%%%%%%%%%%%%%%%%%%%%%%%%%%%

%\newcommand{\fpt}{f_{+}(|{\bf p}|,t)}

\subsection{The Boltzmann equation} %%%%%%%%%%%%%%%%%%%%%%%%%%%%%%%%%%%%%%%

Our goal is to estimate the relic abundance of the right-helicity
neutrinos $\nu_+$
as a function of time. Of particular interest is the contribution
of the right-helicity neutrinos to the effective number of neutrinos,
$\Delta {N}_{\nu}$, at the onset of primeval nucleosynthesis.
To find $\Delta {N}_{\nu}$ one has to determine first the evolution of the
phase
space distribution of the right-helicity neutrinos
$f_+(|\pv_+|,t)$\footnote{The arguments of the distribution
  function $f_{+}(|{\bf p}_{+}|,t)$ indicate that we work in the FRW,
  {\it i.e.}, spatially isotropic and homogeneous cosmology.}.
Then using a primordial nucleosynthesis code
one computes the increase in the primordial $^4{\rm He}$ abundance $\Delta Y_p$
resulting from the non-zero energy densities of $\nu_+$ and
$\overline\nu_-$ at the nucleosynthesis time.
Since $^4{\rm He}$ abundance is a monotonic function of the total energy
density of the universe,
one can translate $\Delta Y_p$ back into $\Delta {N}_{\nu}$, the effective
change of the energy
density in units of that of massless two-component neutrinos. Without using the
NS code one
can approximate $\Delta {N}_{\nu}$ as
\be{nuclsyn}
\Delta {N}_{\nu}\simeq {\rho_{+,{n\leftrightarrow p}}}/
  {\frac{7\pi^2}{240}T_{n\leftrightarrow p}}~,
\ee
where the subscript ${n\leftrightarrow p}$ refers to the freeze-out of the
reactions which
transmute protons and neutrons into each other. The justification  of this
approximation
comes from the fact that $Y_p$ is predominantly determined by the density of
neutrons just
after the ${n\leftrightarrow p}$ freeze-out.

Let us briefly describe the method we have applied for solving the evolution of
the right-helicity
neutrino density. The evolution of the distribution
$f_{+}(|{\pv}_{+}|,t)$ is governed by the Boltzmann equation
\begin{equation}
  \left( \frac{\partial}{\partial t}
         -H|{\pv}_+|\frac{\partial}{\partial |{\pv}_+|} \right) f_+
  = \left( \frac{\partial f_+}{\partial t} \right)_{\rm coll}
  \label{BoltzEq:t}
\end{equation}
with the initial condition
\begin{equation}
  f_+(|{\pv}_{+}|,t_0) = f_+^0(|{\pv}_{+}|)~.
  \label{InitCond:t}
\end{equation}
Instead of time $t$, we will consider the cosmic scale factor $R$ as the
independent variable describing the evolution.
 Eqs.~(\ref{BoltzEq:t}, \ref{InitCond:t}) then transform  into
\begin{equation}
  \left( R \frac{\partial}{\partial R}
         - |{\pv}_+| \frac{\partial}{\partial |{\pv}_+|} \right) f_+
  = \frac{1}{H} \left( \frac{\partial f_+}{\partial R} \right)_{\rm coll}
  \label{BoltzEq:R}
\end{equation}
and
\begin{equation}
  f_+(|{\pv}_{+}|,R(t_0)) = f_+^0(|{\pv}_{+}|)
  \label{InitCond:R}~.
\end{equation}
Since $R$ is defined up to a multiplictive constant
(only the ratio of $R$'s at two different times has physical
meaning), one may choose $R(t_0) = 1$.

For solving  Eqs.~(\ref{BoltzEq:R}, \ref{InitCond:R}) it is important
that { the first-order PDE (\ref{BoltzEq:R}) may be transformed to an
ODE} by a suitable transformation of the variables $|{\pv}_{+}|$ and R.
We use
\begin{equation}
  |{\pv}_{+}| \to \tilde{p}_+ = |{\pv}_{+}|\frac{R}{R_0}~,
  \label{ToPTilde}
\end{equation}
where $R_0$ is arbitrary but fixed.
Then Eq.~(\ref{BoltzEq:R}) may be written as
\begin{equation}
  R\,\frac{\partial}{\partial R} \tilde{f}_+(\tilde{p}_+,R)
  = \frac{1}{H} \left( \frac{\partial \tilde{f}_+}{\partial R}
                \right)_{\rm coll}~,
  \label{BoltzEq:lnR}
\end{equation}
where, according to (\ref{ToPTilde}),
$\tilde{f}_+(\tilde{p}_+,R):=f_+(|{\pv}_+|=
\tilde{p}_+ R_0/R,R)$ and a similar relation holds between the
collision terms $(\partial\tilde f_+/\partial R)_{\rm coll}$
and $(\partial f_+/\partial R)_{\rm coll}$.
Eq.~(\ref{BoltzEq:lnR}) describes the evolution of
$\tilde{f}_+(\tilde{p}_+,R)$ with an increasing $R$ for a fixed value of the
parameter
$\tilde{p}_+$. For noninteracting $\nu_+$'s,
Eq.~(\ref{BoltzEq:lnR}) is solved by
$\tilde{f}_+={const}$, which may be translated into the familiar
evolution (caused by momentum redshift) of the distribution function
of the freely expanding gas of particles,
\begin{equation}
  f_+(|{\pv}_+|,R) = \tilde{f}_+^0(\tilde{p}_+=
                       |{\pv}_+|{R}/R_0)~.
  \label{FreeExpF}
\end{equation}
This equation describes the compression of the distribution function
along the momentum axis as $R$ increases. For ultrarelativistic
particles this compression preserves the shape of the distribution function.
Interactions will in general introduce distortions in the
distribution function because of the momentum-dependent strength of the
coupling of particles with the ambient matter.

As has been already mentioned in Sec.~2.1, we are considering
ultrarelativistic neutrinos and therefore are justified to neglect all
processes involving more than one right-helicity neutrino.
Then, as will be shown in the next subsection, neglecting for a moment a slight
$\tilde{f}_+$-dependence of the Hubble parameter $H$ and the $R-T$
relationship,
used in calculation of both $H$ and the collision term,
the r.h.s. of the Eq.~(\ref{BoltzEq:lnR}) is linear in
the distribution function of right-helicity neutrinos $\tilde{f}_+$,
and consequently (\ref{BoltzEq:lnR}) represents a set of { uncoupled}
differential equations for $\tilde{f}_+(\tilde{p}_+,R)$, one
independent equation for each value of the parameter
$\tilde{p}_+$.\footnote{
  In the general case when one takes into account also processes
  with more than one right-helicitity neutrinos, the r.h.s. is no more
  linear in $\tilde{f}_+$ and one arrives at the set of { coupled}
  ODEs for $\tilde{f}_+$: the evolution of $\tilde{f}_+(\tilde{p}_+,R)$
  and $\tilde{f}_+(\tilde{p}_{+}^\prime,R)$,
 for ${\tilde{p}}_+ \ne \tilde{p}_{+}^\prime$, are not independent.}
In our computations we followed the evolution of the distribution
function at 30 different values ($\sim$ bins) of $\tilde{p}_+$
placed at equal distances in the interval
$0\le \tilde{p}_+/(100 {\rm\ GeV})\le 10$.
We took  into account the weak dependence of $H$ and $T(R)$ on
$\tilde{f}_{+}$ by solving  Eq.~(\ref{BoltzEq:lnR})
iteratively, that is, we evaluated these quantities using the value of
$\tilde{f}_{+}$
obtained in the previous iteration. Because of the  smallness of the correction
to $H$  and $T(R)$ from $\tilde{f}_{+}$ the solution converges rapidly and
we had to use only 3 iterations.

\subsection{The collision term} %%%%%%%%%%%%%%%%%%%%%%%%%%%%%%%%%%%%%%%

Considering only purely fermionic $2\to 2$ and $1\to 3$ processes, we can write
the
collision term on the r.h.s. of (\ref{BoltzEq:lnR}) in the form
\begin{equation}
  \left( \frac{\partial {f}_+}{\partial R} \right)_{\rm coll} =
    (C_{2\to 2}+C_{1\to 3})(1-{f}_+)
  - (C_{2\to 2}^\prime + C_{1\to 3}^\prime) {f}_+~,
  \label{RHS:Cs}
\end{equation}
where  the coefficients $C_I$ represent production and
$C'_I$ destruction of $\nu_+$, with $I = 2\to 2, 1\to 3$.
The explicit expressions for the quantities $C_{2\to 2}$ and $C_{1\to 3}$
are
\begin{eqnarray}
  C_{2\to 2}(|{\pv}_{+}|,R) & = &\sum_{\rm scatt} \frac{1}{2E_+}
    \int d\Pi_a d\Pi_b d\Pi_d (2\pi)^4 \delta^{(4)}(p_a+p_b-p_{+}-p_d)
    \nonumber\\
    & & \quad\quad\quad\quad\quad\quad\quad\times\, S|{\cal
M}_{ab\leftrightarrow {+}d}|^2
    f_a^{\rm FD} f_b^{\rm FD} (1-f_d^{\rm FD})~, \nonumber\\
  C_{1\to 3}(|{\pv}_{+}|,R) & = &\sum_{\rm dec} \frac{1}{2E_+}
    \int d\Pi_f d\Pi_g d\Pi_h (2\pi)^4 \delta^{(4)}(p_f-p_g-p_{+}-p_h)
    \nonumber\\
    & & \quad\quad\quad\quad\quad\quad\quad \times\, S|{\cal
M}_{f\leftrightarrow g{+}h}|^2
    f_f^{\rm FD} (1-f_g^{\rm FD}) (1-f_h^{\rm FD})~,
    \label{Expr:Cs}
\end{eqnarray}
where, as before, $d\Pi_i \equiv d^3p_i/((2\pi)^3 2E_i)$ and $S$ is a
symmetry factor
taking into account identical particles in the initial and/or final states.
Note that these expressions do not include the factors $g_i$ representing
the number of spin degrees of freedom. According to our convention these
factors are included already into the matrix elements squared since
we sum over polarization states of all particles except the right-helicity
neutrino under consideration. We have also used the well justified assumption
that charged leptons, quarks and left-helicity neutrinos have thermal FD
distributions.

The expressions for  $C'_I$ can be obtained from (\ref{Expr:Cs}) by the
replacement
\begin{equation}
  f_i^{\rm FD} \leftrightarrow (1-f_i^{\rm FD})~.
  \label{CToCPrime}
\end{equation}
Using (\ref{Expr:Cs}) and (\ref{CToCPrime}) one easily finds that the
primed and unprimed coefficients are related to each other by
\begin{equation}
  C_I^\prime = e^{E_{+}/T} C_I~,
  \label{CToCPrime1}
\end{equation}
where we have assumed that all the particles except the right-helicity
neutrino are at a common temperature $T$ and the chemical potentials of all the
particles
can be neglected. With Eq.~(\ref{CToCPrime1}) we are able to rewrite
(\ref{RHS:Cs})
in the more compact form
\be{compact}
 \left( \frac{\partial {f}_+}{\partial R}\right) _{\rm coll}=(C_{2\to 2} +
C_{1\to 3})
\left[
1-\frac{f_+}{f_+^{\rm FD}}\right]~,
\ee
where $f_+^{\rm FD}=[\exp(E_+/T)+1]^{-1}$.

Let us introduce the total production rates per unit volume:
\begin{eqnarray}
  \Gamma_{2\to 2}(R) & = &\sum_{\rm scatt}
    \int d\Pi_a d\Pi_b d\Pi_+ d\Pi_d (2\pi)^4 \delta^{(4)}(p_a+p_b-p_{+}-p_d)
    \nonumber\\
    & & \quad\quad\quad\quad\quad\quad\quad \times\, S|{\cal
M}_{ab\leftrightarrow {+}d}|^2
    f_a^{\rm FD} f_b^{\rm FD} (1-f_d^{\rm FD})~, \nonumber\\
  \Gamma_{1\to 3}(R) & = &\sum_{\rm dec}
    \int d\Pi_f d\Pi_g d\Pi_+ d\Pi_h (2\pi)^4 \delta^{(4)}(p_f-p_g-p_{+}-p_h)
    \nonumber\\
    & & \quad\quad\quad\quad\quad\quad\quad \times\, S|{\cal
M}_{f\leftrightarrow g{+}h}|^2
    f_f^{\rm FD} (1-f_g^{\rm FD}) (1-f_h^{\rm FD})~.
    \label{Expr:Gammas}
\end{eqnarray}
Note that the expressions given above do not include the blocking factors
$(1-f_+)$. Also, the dependence of $\Gamma_I$'s on $R$ arise through the
distribution functions of ambient particles.
Comparing Eqs.~(\ref{Expr:Cs}) and (\ref{Expr:Gammas}) one easily sees that
\begin{equation}
  \Gamma_I(R) = \int\frac{d^3p_+}{(2\pi)^3}\,C_I(|{\pv}_+|,R)~,
\end{equation}
or equivalently
\begin{equation}
  C_I(|{\pv}_+|,R) = \frac{d}{d(|{\pv}_+|^3/(6\pi^2))}\;\Gamma_I(R)~.
  \label{GammaToC}
\end{equation}
{}From this expression we see that the quantities $C_I$, which are of the
dimension of mass, are
the production rates of $\nu_+$'s per unit volume per unit interval of
$|{\pv}_+|^3/(6\pi^2)$ around $|{\pv}_+|$.

Eq.~(\ref{GammaToC}) suggests also a general method for calculating
$C_I$'s. One first generates by a Monte Carlo (MC) method a sample of
unweighted  events, together with their common weight, such that the total
integral
over the initial and final state phase space yields $\Gamma_I$. Having
this sample, the evaluation of the derivative in (\ref{GammaToC})
essentially reduces to building the histogram the number of unweighted 
events vs. $|{\pv}_+|^3/(6\pi^2)$.
Although this method is very general (applicable both to scatterings and
decays) and with a clear physical meaning, it yields a good accuracy only
if one generates the initial sample of weighted events
such that there are more weighted events in the regions of the phase space
where the integrand in $\Gamma_I$ is larger.
This requires some knowledge of the behaviour of the integrand
and an ability to build a MC event generator with the required
distribution of events in the phase space.

In the case of $2\to 2$ scatterings there exists also another method
of calculation of $C_{2\to 2}$. This method follows directly from
(\ref{Expr:Cs})
and is based on the T-invariance of interactions allowing to rewrite
$C_{2\to 2}$ in a form where $\nu_+$ is an initial state particle. After
this transformation the calculation of the corresponding integral
proceeds in a usual way.
In order to check our results we have  determined $C_{2\to 2}$ with both
methods.

\subsection{Results} %%%%%%%%%%%%%%%%%%%%%%%%%%%%%%%%%%%%%%%%%%%%%%%%%%%%

The total rates $C(|{\bf p}_+|,R)=C_{2\to 2}(|{\bf p}_+|,R)+C_{1\to
3}(|{\bf p}_+|,R)$ for the right-helicity tau neutrino
production for a number of fixed
momenta are shown in Fig.~2. One can clearly see the effect of the s-channel
pole, indicating that the rate is dominated by fermionic s-channel
processes. Small momentum states pass through the pole at high
temperatures, large at low temperatures. Another feature is the increase
of the total rate for very small momentum $\nu_+$'s at
$0.2~{\rm GeV} \lsim T \lsim 2~{\rm GeV}$.
This enhancement is due to both scatterings
($u\overline{d}\to{\nu_+\tau^+}$,
$c\overline{s}\to{\nu_+\tau^+}$) and the decays of the tau lepton. Since at
these temperatures
the thermally averaged center of mass energy of colliding particles is of the
same
order of magnitude as the mass of the tau lepton, the neutrinos (and $\tau$'s)
are preferentially produced in the
small-momentum states, for which the probability of the creation
of a right-helicity neutrino (spin-flip) is larger. For similar reasons
the contribution of $\tau$-decays to the total rate can be as large as
$\sim 30$\%. For right-helicity muon neutrinos one expects the analogous
mechanisms to be effective at temperatures below 100 MeV.

Because different momentum states have different interaction rates,
this causes a distortion of the momentum distribution of $\nu_+$'s  relative to
the equilibrium distribution. This effect is demonstrated in Fig.~3.
Here we have parametrised the distribution function of $\nu_+$ for fixed R,
$f_+(|{\bf p_+}|)$, introducing
a momentum-dependent effective temperature $T_{\rm eff}(|{\bf p_+}|)$ through
$f_+(|{\bf p_+}|) = [\exp(E_+/T_{\rm eff}(|{\bf p_+}|))+1]^{-1}$.
In Fig.~3  the ratio $T_{\rm eff}/T$ for a tau neutrino of mass 10~keV is shown
at
$T = 30$, 3 and 0.3 GeV for two extreme initial conditions at a high
temperature
$T=100$ GeV: ({\it i}) $f_{+}=f_{-}=f^{\rm FD}_+$, {\it i.e.} complete
equilibrium, and
({\it ii}) $f_{+}=0$. One sees that only the lowest momentum right-helicity
tau neutrinos of this mass will come into full equilibrium at the pole:
$T_{\rm eff}$ depends on the initial condition even at $T=0.3$ GeV for
$|{\bf p}_+|/T \gsim 1$.
The rise of $T_{\rm eff}$ at the small-momentum end of the spectrum is because
these states,
interacting more strongly, are kept in good thermal contact with ambient
matter even when higher momentum states are (already) decoupled. This kind
of the distortion of the distribution function becomes more
prominent as the mass of the neutrino increases. As
mentioned before, the regime of free expansion reveals itself in
the shape-preserving
evolution of the distribution function. Another remarkable feature is
the rise of $T_{\rm eff}$ at higher momenta clearly seen for the curve
corresponding
to $T=30$~GeV and the zero initial condition $f_+=0$. This rise is there
despite the fact
the total production rate $C(|{\bf p}_+|,R)$ is a monotonically
(at least for higher momenta) dereasing function of $|{\bf p}_+|$.
The explanation can be found from Eq.~(\ref{compact}):
if $f_+/f_+^{\rm FD}$ is much smaller than unity, the destruction
of $\nu_+$'s can be neglected. The regime of well-out-of-equilibrium
production of $\nu_+$ is effectively maintained only while
both the interaction rate $C(|{\rm p}_+|,R)$ and
the ratio $f_+/f_+^{\rm FD}$ are sufficiently small. These conditions are more
easily fulfilled by the higher momentum neutrinos at the beginning of their
evolution from the zero initial condition.
This out-of-equilibrium production results in the distribution function
decreasing with $|{\bf p}_+|$ more slowly than the equilibrium distribution
function, or, equivalently, an $T_{\rm eff}$ rising with $|{\bf p}_+|$.
In contrast with the above discussed rise
of $T_{\rm eff}$ at lower momenta, the later feature becomes more apparent as
the mass of
the neutrino decreses.
We find that spectral
distortion changes the energy desity of $\nu_+$'s (which were in equilibrium at
temperatures above the weak interaction pole) at $\sim 1$ MeV typically by only
a few percent.
The major difference between the actual energy density and the naive estimate
$\rho_R\simeq 0.044\: \rho_L$ is due to the relatively late
decoupling caused by the pole.

Once the initial condition is fixed, we may follow the evolution of
the energy density of $\nu_+$.
In Fig.~4 we show the evolution of the right-helicity neutrino energy density
for $m_{\nu_\tau}=1,~6~{\rm and}~20\keV$ for two different initial conditions.
If $m_{\nu_\tau}=1\keV$, spin-flip interactions are too weak to affect the
evolution, and the resulting relic abundance at nucleosynthesis is, to
a high accuracy, what one would naively expect. However, if
$m_{\nu_\tau}=20\keV$,
spin-flip interactions at the pole are strong enough to equilibrate
$\nu_+$ even if initially $f_{+}=0$.

The right-helicity neutrino
energy density at about nucleosynthesis time in units of left-helicity
neutrino energy density is shown in Fig.~5.
We see that almost full equilibration of $\nu_+$
is obtained at some temperature below 100 GeV if $m_{\nu_\tau}\gsim 10\keV$.
We have assumed here that there is no entropy production at QCD phase
transition
so that the only effect is the dilution of the right-helicity neutrino
densities by the appropriate ratios of the effective degrees of freedom
before and after the phase transition.  We have computed $\Delta {N}_{\nu}$
assuming also that below $T_{\rm QCD}$ all interactions can be ignored.
While this is a very good approximation
for smaller masses ($m_{\nu_\tau} \lsim 30$ keV),
it has been found \cite{kimmo,FieldsKimmoOliivi} that for higher masses
out-of-equilibrium scatterings and the decay
$\pi^0\to\nu_{+}\overline{\nu}_{+}$
(and the decay $\pi^{+}\to\mu^{+}\nu_{+}$ in the case of muon
neutrinos) produce considerable amounts of $\nu_+$. Using the
results of \cite{FieldsKimmoOliivi} we estimate that the non-equilibrium
reactions occuring below $T_{\rm QCD}$ give an additional contribution
to $\Delta {N}_{\nu}$, to be added to our result shown in Fig.~5,
which is  approximately
0.003 for $m_{\nu_\tau} = 30$ keV,
0.012 for $m_{\nu_\tau} = 60$ keV,
0.03 for $m_{\nu_\tau} = 100$ keV and
0.15 for $m_{\nu_\tau} = 200$ keV, adopting $T_{\rm QCD}=150$~MeV.
Given the inaccuracy of the determination of $N_{\nu}$ from the
observational data, at present the pole effect cannot be used to derive an
upper bound on the masses of Dirac neutrinos.

We have assumed no other interactions than those present in the Standard Model.
Any non-standard interaction above the electroweak scale would affect the
initial
energy density of the right-handed neutrinos. However, according to our
results,
if the mass of neutrino is more than about 10 keV,
nucleosynthesis is not  sensitive
to such interactions as their effect will be washed out by equilibration
at the electroweak pole region. Because this equilibration is inefficient
below neutrino masses in the 1 keV range, the cosmological mass limit
of stable Dirac neutrinos is not modified. The increase is at most about
4\% as expected naively, depending on the unknown initial density at high
temperatures.
\vskip1cm
\noindent
{\Large\bf Acknowledgements}
\vskip0.2cm
The work was supported by Academy of Finland.
One of us (P.K.) would like to thank the Finnish Cultural Foundation and
its Pohjois-Karjala Foundation for financial support,
and another (H.U.) the Nordic Council of Ministers and CIMO for
grants.
We also wish to thank Kimmo Kainulainen for useful discussions.

\newpage

%%%%%%%%%%%%%%%%%%%%%%%%%%%%%%%%%%%%%%%%%%%%%%%%%%%%%%%%%%%%%%%%%%%%%%%%%%%%%%

%%%%%%%%%%%%%%%%%%%%%%%%%%%%%%%%%%%%%%%%%%%%%%%%%%%%%%%%%%%%%%%%%%%%%%%%%%%%%%%
%  TABLE
%%%%%%%%%%%%%%%%%%%%%%%%%%%%%%%%%%%%%%%%%%%%%%%%%%%%%%%%%%%%%%%%%%%%%%%%%%%%%%%
\newpage
{\begin{table*}
\centering
\caption[t2]{Processes, up to crossings, which produce a right-helicity tau
neutrino $\nu_+$.}
\vskip0.5cm
\begin{tabular}{|c|c|c|c|}\hline
${\rm fermionic~ 2\to 2}$ &$ {\rm bosonic~ 2\to 2}$&${\rm 3-body~ decays}$&
${\rm 2-body~ decays}$\\
\hline
$ \nu_{\tau} \nu_{\tau} \to \nu_{+} \nu_{\tau} $ &
  $ \nu_{\tau} W^-  \to  \nu_{+} W^- $ &
  $ \tau^- \to \nu_{+} l_j^- \overline\nu_j\,, $ &
  $ W^+  \to  \tau^+ \nu_{+} $  \\
& & $(l_j^- \overline\nu_j = e^- \overline\nu_e,\:\mu^-\overline\nu_\mu)$ & \\
$ \nu_{\tau} \nu_j \to \nu_{+} \nu_j\,,~(j=e,\mu ) $ &
  $ \nu_{\tau} Z  \to  \nu_{+} Z  $ &
  $ \tau^- \to  \nu_{+} d_n \overline{u}_m\,, $ &
  $ Z \to \nu_{+} \overline{\nu}_{\tau} $  \\
& & $ (d_n \overline{u}_m = d \overline{u},\:
                            s \overline{u},\:
                            s \overline{c},\:
                            d \overline{c})$ & \\
$ \nu_{\tau} \tau^- \to \nu_{+} \tau^-$ &
  $ \nu_{\tau} H  \to  \nu_{+} H $ &
  $ t \to d_n \tau^+ \nu_{+}\,,$ &
  $ H \to \nu_{+} \overline{\nu}_{\tau} $ \\
& & $ (d_n = b,\:s,\:d) $ & \\
$ \nu_{\tau} f_{\rm ch} \to \nu_{+} f_{\rm ch}\,,~
                                   (f_{\rm ch}\ne \tau^-) $ &
  $ \tau^- Z  \to  \nu_{+} W^- $ &
  $ \overline{b} \to \overline{u}_m \tau^+ \nu_{+}\,, $ & \\
& & $ (\overline{u}_m = \overline{c},\:\overline{u}) $ & \\
$ \tau^- \nu_j  \to  \nu_{+} l_j^-\,,~ (j=e,\mu) $ &
  $ \tau^- H \to \nu_{+} W^- $ & & \\
& & & \\
$ \tau^-u_m\to\nu_+d_n$ &
  $\tau^- \gamma \to \nu_{+} W^- $ & & \\
& & & \\
\hline
\end{tabular}
\label{t2}
\end{table*}
\null

%%%%%%%%%%%%%%%%%%%%%%%%%%%%%%%%%%%%%%%%%%%%%%%%%%%%%%%%%%%%%%%%%%%%%%%%%%%%%%%
%  FIGURE CAPTIONS
%%%%%%%%%%%%%%%%%%%%%%%%%%%%%%%%%%%%%%%%%%%%%%%%%%%%%%%%%%%%%%%%%%%%%%%%%%%%%%%
\newpage\noindent
{\Large\bf Figure captions}
\vskip2cm

\noindent{\bf Figure 1.}
Thermally averaged production rates per one right-helicity tau neutrino
$\nu_+$ of mass 20~keV for
the s-channel process $u\bar d\to\nu_+\tau^+$,
the t-channel process $\tau^-u\to \nu_+ d$ and
the bosonic process $\tau^-\gamma\to\nu_+ W^-$.
The Hubble expansion rate is also shown for comparision.
\vskip0.5cm

\noindent{\bf Figure 2.}
The total production rates (dashed curves) $C(|{\bf p}_+|,R)$ of right-helicity
tau neutrinos as functions of the temperature $T$. From  top to   bottom,
the rates are given for  $\tilde p_+/100~{\rm
GeV} = 0.33, 1, 2 ,4, 6, 8$ and $10$ . For comparison
the evolution of the Hubble parameter $H$ is also given (solid curve).
\vskip0.5cm

\noindent{\bf Figure 3.}
The ratio of the momentum-dependent effective temperature $T_{\rm eff}$
of the right-helicity tau neutrinos of mass 10~keV to the
plasma temperature $T$
at $T=30$~GeV, 3~GeV and 0.3~GeV. In each case the solid curve corresponds to
the equilibrium initial condition $f_+ = f_+^{\rm FD}$, and the dashed
curve to the zero initial condition $f_+ = 0$ at $T=100$~GeV.
\vskip0.5cm

\noindent{\bf Figure 4.}
The evolution of $\rho_+/\rho_+^{\rm FD}$ of tau neutrinos as a function of
the temperature $T$ for
$m_{\nu_{\tau}}=20$~keV (solid curve),
6~keV (dot-dashed curve) and 1~keV (dashed curve).
The evolution is given for two different initial conditions:
$f_+=f_+^{\rm FD}$ and $f_+=0$
at $T=100$~GeV.
\vskip0.5cm

\noindent{\bf Figure 5.}
The energy density of right-helicity tau neutrinos $\nu_+$ in units of the
effective number of two-component
neutrino species, $\Delta {N}_{\nu}$, as a function of the neutrino mass.
$\Delta {N}_{\nu}$ is calculated at $T=3$~MeV (solid curves),
1~MeV (dashed curves)and 0.3~MeV (dot-dashed curves).
For $m_{\nu_{\tau}} < 20$~keV the two sets of curves differ by
initial conditions at $T=100$~GeV.
%%%%%%%%%%%%%%%%%%%%%%%%%%  FIGURES %%%%%%%%%%%%%%%%%%%%%%%%%%%%%%%%%%%%
\newpage
\begin{figure}
\leavevmode
\centering
\vspace*{120mm}
\includegraphics{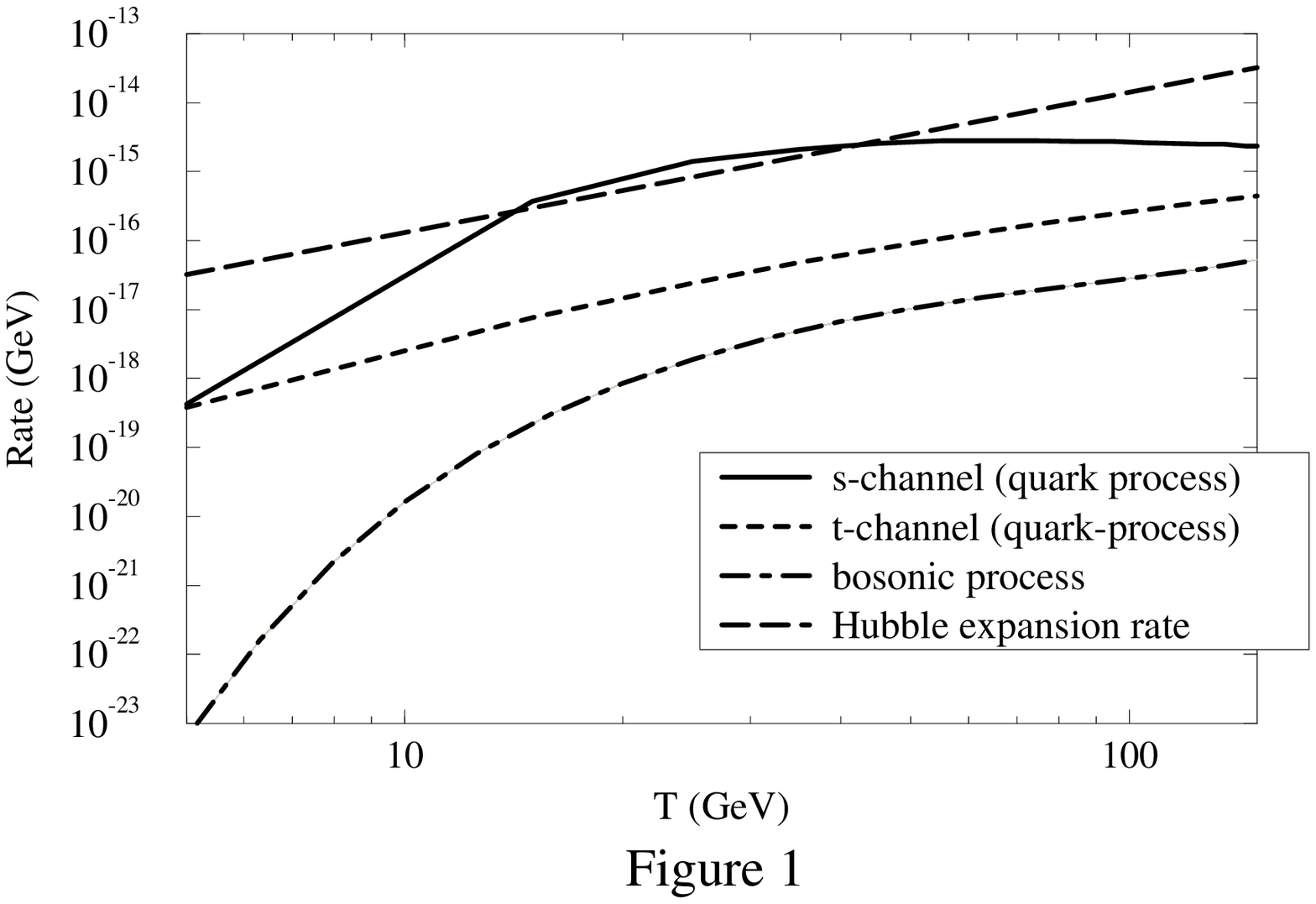}
\label{kuva1}
\end{figure}
%%%%%%%%%%%%%%%%%%%%%%%%%%%%%%%%%%%%%%%%%%%%%%%%%%%%%%%%%%%%%%%%%%%%%%%
\begin{figure}
\leavevmode
\centering
\vspace*{80mm}
\includegraphics{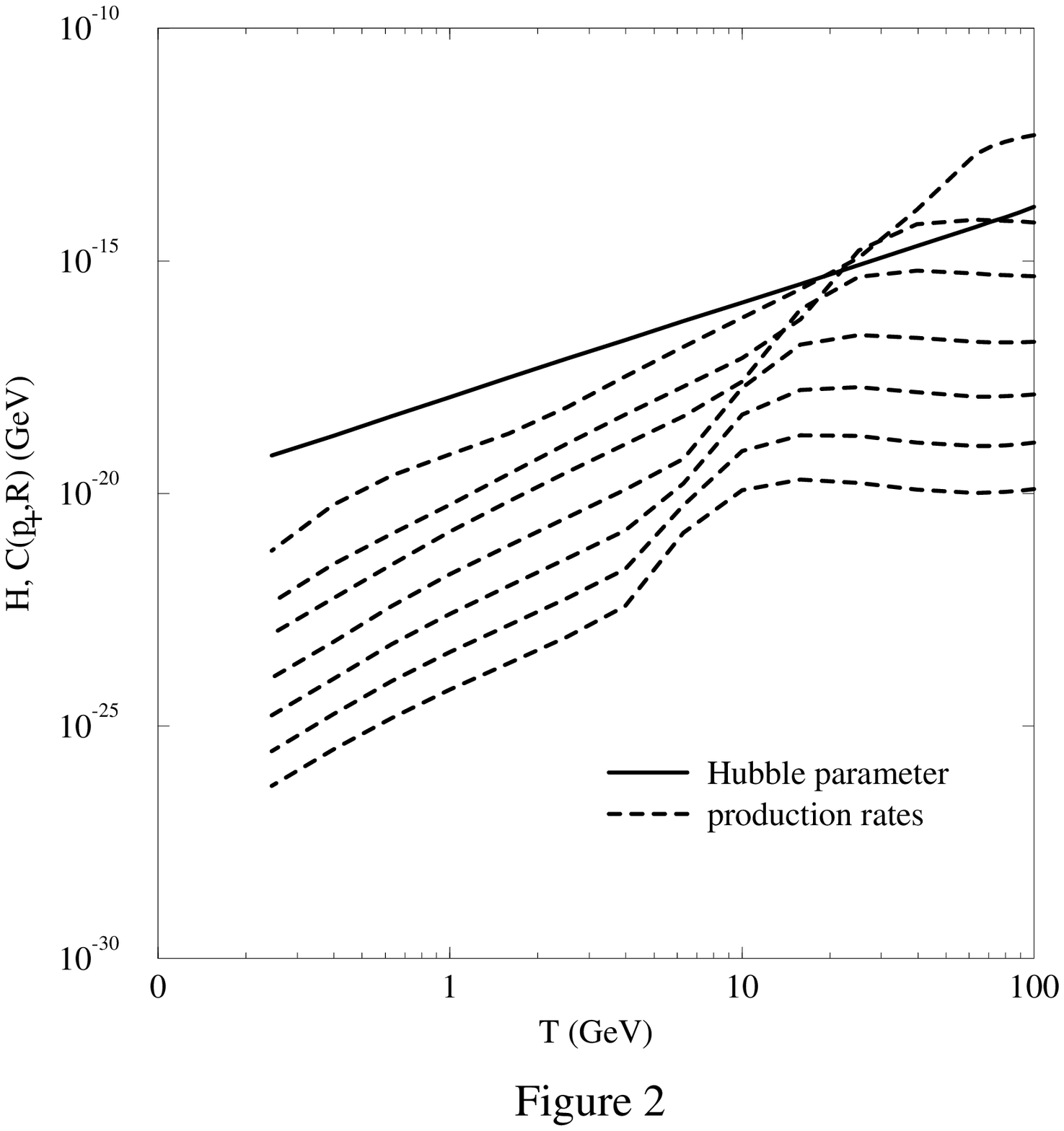}
\label{kuva2}
\end{figure}
%%%%%%%%%%%%%%%%%%%%%%%%%%%%%%%%%%%%%%%%%%%%%%%%%%%%%%%%%%%%%%%%%%%%%%%
\newpage
\begin{figure}
\leavevmode
\centering
\vspace*{80mm}
\includegraphics{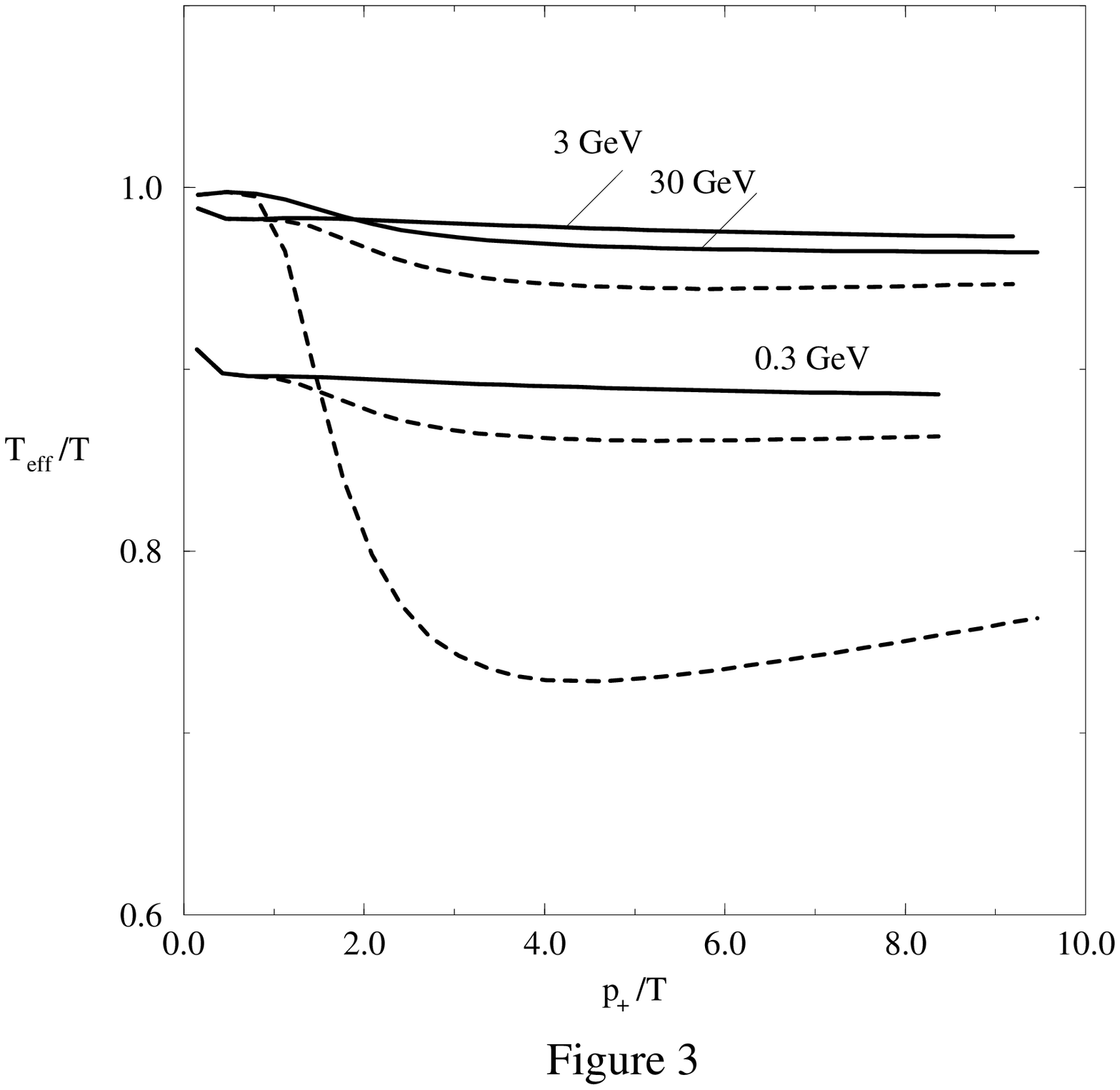}
\label{kuva3}
\end{figure}
%%%%%%%%%%%%%%%%%%%%%%%%%%%%%%%%%%%%%%%%%%%%%%%%%%%%%%%%%%%%%%%%%%%%%%%
\newpage
\begin{figure}
\leavevmode
\centering
\vspace*{80mm}
\includegraphics{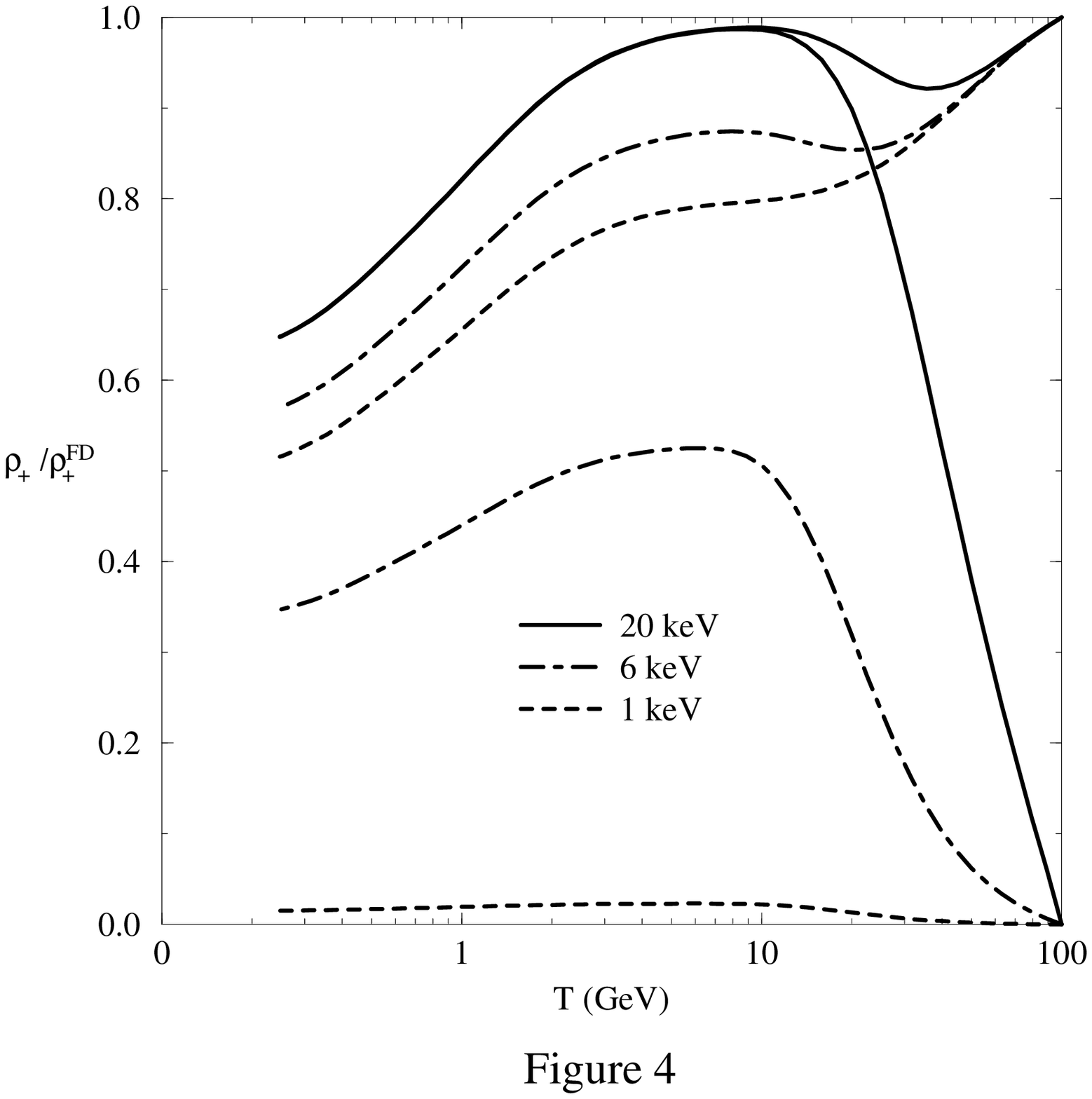}
\label{kuva4}
\end{figure}
\newpage
%%%%%%%%%%%%%%%%%%%%%%%%%%%%%%%%%%%%%%%%%%%%%%%%%%%%%%%%%%%%%%%%%%%%%%%
\begin{figure}
\leavevmode
\centering
\vspace*{80mm}
\includegraphics{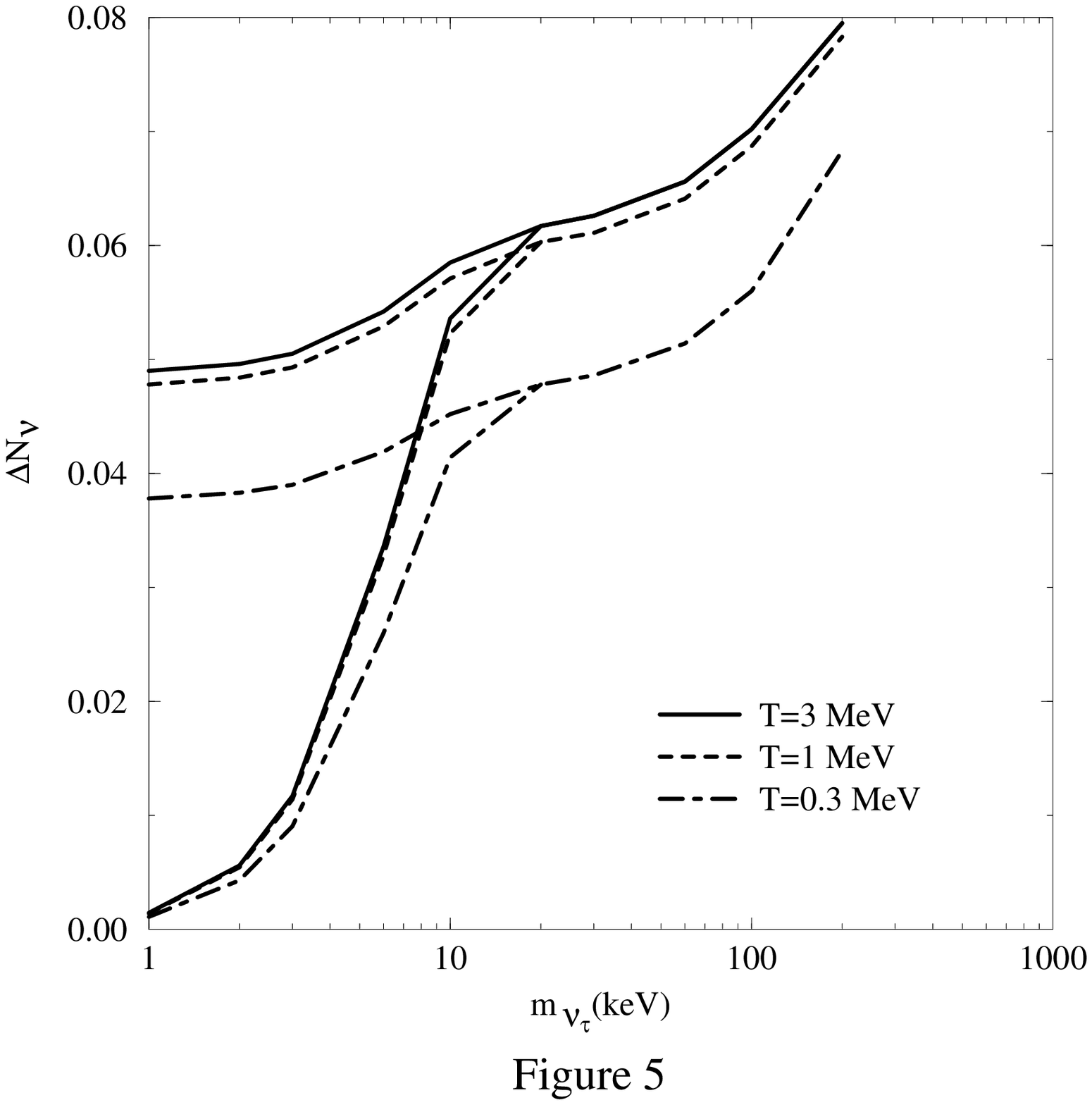}
\label{kuva5}
\end{figure}
%%%%%%%%%%%%%%%%%%%%%%%%%%%%%%%%%%%%%%%%%%%%%%%%%%%%%%%%%%%%%%%%%%%%%%%
\end{document}